\documentclass[journal=apchd5,manuscript=article,layout=traditional]{achemso}

\usepackage{graphicx}
\usepackage[intlimits]{amsmath}
\usepackage{amssymb}
\usepackage{amsthm}
\usepackage{booktabs}
\usepackage{multirow}
\usepackage{natbib}
\usepackage{geometry}
\usepackage{color}
\usepackage{bm}

\author{R. Carmina Monreal}
\affiliation{Departamento de F\'{\i}sica Te\'orica de la Materia Condensada C5 and Condensed Matter Physics Center (IFIMAC), Universidad Aut\'onoma de Madrid, E-28049 Madrid, Spain}
\author{Tomasz J. Antosiewicz}
\affiliation{Centre of New Technologies, University of Warsaw, Banacha 2c, 02-097 Warsaw, Poland}
\alsoaffiliation{Department of Applied Physics and Gothenburg Physics Centre, Chalmers University of Technology, SE-412 96 G\"oteborg, Sweden}
\author{S. Peter Apell}
\email{peter.apell@chalmers.se}
\affiliation{Department of Applied Physics and Gothenburg Physics Centre, Chalmers University of Technology, SE-412 96 G\"oteborg, Sweden}

\title{Diffuse Surface Scattering in the Plasmonic Resonances of Ultra-Low Electron Density Nanospheres}


\begin{document}

\begin{abstract}
Localized surface plasmon resonances (LSPRs) have recently been identified in extremely diluted electron systems obtained by doping semiconductor quantum dots. 
Here we investigate the role that different surface effects, namely electronic spill-out and diffuse surface scattering,
play in the optical properties of these ultra-low electron density nanosystems.
Diffuse scattering originates from imperfections or roughness at a microscopic scale on the surface.
Using an electromagnetic theory 
that describes this mechanism in conjunction with a dielectric function including the quantum size effect,
we find that the LSPRs show an oscillatory behavior both in position and width 
for large particles and a strong blueshift in energy and an increased width for smaller radii, consistent with recent experimental results for photodoped ZnO nanocrystals. 
We thus show that the commonly ignored process of diffuse surface scattering is a more important mechanism affecting the plasmonic properties of 
ultra-low electron density nanoparticles than the spill-out effect.

\includegraphics{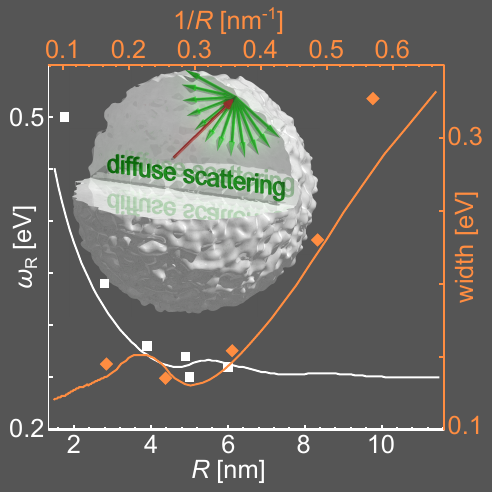}

\end{abstract}

\maketitle

\newpage

Surface plasmons in metals have been studied for a long time \cite{PhysRev_106_874_ritchie, surface_plasmons_reather, liebsch_book_1997}. 
However, in the last few years, the behavior of plasmon resonances in nanosystems has become a hot topic due to numerous applications\cite{plasmonics_maier}, having been made possible by considerable progress in both calculating and measuring the optical properties of nanoparticles.
A particularly interesting example concerns the energy shift experienced by the localized surface plasmon resonances (LSPRs) of nanospheres
as a function of the size, as well as the corresponding modification in the plasmon width 
\cite{Nature_483_421_scholl, nanophotonics_2_131_raza, PNAS_107_14530_peng, NL_9_3463_baida, NL_12_429_townsend, JPCL_1_2922_lerme}.

Recently, it was discovered that doped semiconductor quantum dots 
with a high concentration of carriers in the conduction band show, when illuminated, absorption resonances attributed to the excitation of surface plasmons
\cite{NMat_10_361_luther, NL_11_4706_buonsanti, PNAS_109_8834_naik, JPCL_4_3024_faucheux, JPCL_5_976_faucheaux, ACSNano_8_1065_schimpf}. 
Not only are doped semiconductor nanocrystals emerging as a new class of plasmonic structures for their remarkable properties of tuning and switching on/off
the plasmon energy \cite{JPCL_4_3024_faucheux, JPCL_5_976_faucheaux, but also these crystals are of fundamental interest as quantum objects since their size can be of the order of the Fermi wave length of the electrons.}
The density of electrons in such nanocrystals is low, typically two to three orders
of magnitude lower than in typical metals. In terms of the classic density parameter $r_s=(\frac{3}{4 \pi n_e})^\frac{1}{3}$, 
$n_e$ being the carrier density, which in units of the Bohr radius
 is 2-6 for common metals, $r_s$ in doped semiconductors is on the order of 25. Such a diluted electron gas has its absorption band  in the infrared rather than 
in the visible part of the electromagnetic spectrum.

In previous publications \cite{NJP_15_083044_carmina, OpEx_22_24994_carmina} we analyzed how the so-called surface spill-out effect
modifies the position and width of plasmonic resonances in 
Ag and Au nanospheres of radii smaller than 10 nm embedded in different dielectric matrices. 
This effect originates from the fact that the electronic density
of typical high electron density metals does not have a sharp discontinuity at the geometrical surface, as assumed by the classical theory. 
Rather, it is a continuous function because the surface potential barrier confining the electrons is finite and smooth. 
These studies omitted another surface mechanism which affects local surface plasmon resonances in any geometry, and is associated with 
the existence of imperfections or roughness at a microscopic scale on the surface. As a consequence, some of the metal electrons arriving at the surface are reflected back in any direction 
\cite{JPhysFr_38_863_flores, PS_22_155_monreal, JPhysFr_43_901_monreal}.
 The origin of the phenomenon can be better understood
in the case of planar surfaces. For an ideally flat surface, the translational invariance of space in the direction parallel to the surface 
implies conservation of the parallel momentum and, therefore, conservation of the parallel component of the electric current
(specular scattering). However, this  restriction is lifted for non-ideal surfaces, where it is possible to have 
diffuse scattering, and, consequently, fluctuations in the parallel current. 
This mechanism affects any physical phenomenon taking place at a surface 
and, in particular, will lead to energy broadening of the LSPRs since it breaks coherence of single scattering events.
 For planar surfaces or spheres with typical metallic electronic densities, the spill-out mechanism dominates over the diffuse surface
scattering one at the frequencies of the surface plasmon modes, this being the reason why the second mechanism is generally neglected \cite{ProgSurfSci_12_287_feibelman}.
However, imperfections in the shape and morphology of ultra small particles are frequent and constitute an important source for diffuse surface scattering, 
particularly for the low-density electron systems that we will study in this work. Similar behavior can be produced by adsorbed molecules \cite{SurfSci_281_153_persson, JPCC_118_28075_mogensen}.

The purpose of this article is to investigate the role that spill-out and diffuse surface scattering  play in the optical properties of ultra-low electron density nanosystems. Using an electromagnetic theory 
that describes diffuse surface scattering in conjunction with a dielectric function that includes the effect of quantization due to the small size of the particles, we find that the LSPRs show an oscillatory behavior both in position and width 
for particles larger than ca. 3 nm in radius and a strong blueshift in energy and an increased width for smaller radii, consistent with recent experimental results \cite{ACSNano_8_1065_schimpf}. 
We thus show that the commonly ignored process of diffuse surface scattering is an important mechanism affecting the plasmonic properties of these systems.
Based on our theory, we can also extract an effective length characterizing diffuse surface scattering 
which can be included in an effective dielectric function and used in simpler models.

Surface scattering mechanisms affecting the electromagnetic response of metal surfaces can be described by means of two 
effective complex lengths, $d_{\perp}(\omega)$ and $d_{\parallel}(\omega)$, where $\omega$ is the frequency. These are surface response functions 
associated with changes across the surface of the normal
component of the electric field vector and of the parallel component of the displacement vector with respect to their classical
counterparts, respectively \cite{ProgSurfSci_12_287_feibelman}. 
The length $d_{\perp}(\omega)$ can be related to the electronic charge density, $\delta\rho$, induced at the surface by any external perturbation and describes the spill-out effect, while diffuse surface scattering, which provokes fluctuations in the current parallel to the metal surface, 
is described by $d_{\parallel}(\omega)$ ($d_{\parallel}(\omega)=0$ for specular scattering) \cite{ProgSurfSci_12_287_feibelman}. 
For the case of a  sphere 
of radius $R$, the corresponding lengths are denoted as $d_r(\omega, R)$ and $d_{\theta}(\omega, R)$ and,
by considering the mentioned changes in $E_r$ and $D_{\theta}$ across the sphere surface respectively, it was shown 
\cite{PS_26_113_apell, PRL_50_1316_apell} that, in the quasi-static limit $\frac{\omega}{c}R \ll 1$, 
the polarizability of the sphere  can be written as
\begin{equation}
\alpha(\omega)=R^3 \frac{\left(\epsilon(\omega)-\epsilon_m\right)\left(1-\frac{d_r(\omega, R)}{R}\right)+2\frac{d_{\theta}(\omega, R)}{R}}
{\epsilon(\omega)+2\epsilon_m+2\left(\epsilon(\omega)-\epsilon_m\right) \frac{d_r(\omega, R)}{R}+2 \frac{d_{\theta}(\omega, R)}{R}}.
\label{alpha}
\end{equation}

Then, in the same quasi-static limit,  the optical absorption cross section can be calculated as
\begin{equation}
\sigma(\omega)=4 \pi \frac{\omega}{c} \sqrt{\epsilon_m}\, \mathrm{Im}[ \alpha(\omega)],
\label{sigma-abs}
\end{equation}
where $c$ is the speed of light.  
In eqs. (\ref{alpha}) and (\ref{sigma-abs}), $\epsilon(\omega)$ is the classical, 
local permittivity of the metal and $\epsilon_m$ is a frequency independent permittivity of the surrounding medium. 
Note that if surface effects are neglected  ($d_r=0$ and $d_{\theta}=0$ in eq. (\ref{alpha})),
the classical results of the Mie theory are recovered.

The role that these two lengths play in the effective relaxation time of the surface plasmons in small spheres with typical 
electron densities characterized
by $r_s=$2, 3 and 4, was analyzed in Ref. \cite{SSComm_52_971_apell}. At these densities the  spill-out 
contribution to the surface plasmon relaxation time is much larger than the diffuse surface scattering contribution, same as for planar surfaces.
However, as the electron density decreases, the spill-out contribution decreases quickly while the diffuse surface scattering 
one remains nearly constant (we refer the reader to Fig. 2 of Ref. \cite{SSComm_52_971_apell}). 
Therefore, one expects the relative importance of both contributions to be reversed for low enough
electron densities.

These contributions can be estimated rather easily from eq. (\ref{alpha}) and the results of Ref. \cite{SSComm_52_971_apell} (see Supporting Information, SI, for a detailed derivation). Assuming $\epsilon_m=1$ and $\epsilon(\omega)=1-(\frac{\omega_p}{\omega})^2$, $\omega_p$ being the bulk plasma frequency,
we have the classical Mie LSPR frequency of the sphere $\omega_{s}^{cl}=\omega_p/\sqrt{3}$.
From eqs. (\ref{alpha}) and (\ref{sigma-abs}) one can obtain its full-width at half-maximum (see eq. (S6)).
Substituting $d_r$ and $d_{\theta}$ by their counterparts for a planar surface ($d_{\perp}(\omega)$ and $d_{\parallel}(\omega)$, respectively),  
a good approximation for large $R$, we find that for a perfectly diffuse surface reflecting electrons at random, the ratio
of spill-out to diffuse surface scattering contributions can be estimated as

\begin{equation}
\frac {\Gamma_{spill-out}}{\Gamma_{diffuse}} \simeq 2.33 \sqrt{\frac{m^*}{r_s}} 
\frac {\mathrm{Im}[-d_{\perp}(\omega_{s}^{cl})]}{a_0},
\label{gamma-r}
\end{equation}
which is cast in terms of the effective electron mass $m^*$ (in units of the electron mass $m_e$) and the one-electron radius $r_s$ (in units of the Bohr radius $a_0$), with $a_0=\frac{\hbar^2}{m_e e^2}$.

Equation~(\ref{gamma-r}) shows that the spill-out contribution decreases for effective masses smaller than the electron mass,
which is the case for the conduction electrons in doped semiconductors.
Also, the ratio of the contributions of the two mechanisms to the plasmonic width decreases with increasing $r_s$ as
$ \mathrm{Im}[-d_{\perp}(\omega_{s}^{cl})]/ \sqrt{r_s}$. Taking the values of $\mathrm{Im}[-d_{\perp}(\omega_s^{cl})]$ for $r_s=2$ - 5 reported in 
Ref. \cite{PRB_36_7378_liebsch} and $m^{*}=1$, one obtains values of 
$\frac {\Gamma_{spill-out}}{\Gamma_{diffuse}} $ decreasing from 2.5 for $r_s=2$, to 0.5 for $r_s=5$. 
This estimate, as well as a simple analysis in the SI, show that the effects of non-specular scattering become more important as the electron density is lowered and will largely overcome these of electronic spill-out for $r_s \simeq 25$.

Having demonstrated the importance of accounting for diffuse surface scattering, we now proceed to analyze these effects in a detailed manner.
A microscopic description of this effect is, however, beyond the scope of the present work. Instead we use a
phenomenological theory first designed for planar surfaces \cite{JPhysFr_38_863_flores, PS_22_155_monreal, JPhysFr_43_901_monreal}
and then extended to spheres \cite{PRB_32_7878_deAndres}. 
The theory embeds the real sphere in an infinite, fictitious medium having exactly the same dielectric functions. Then, a constitutive relation giving the
polarization $\mathbf{P}_f(\mathbf{r}, \omega)$ due to the free charges inside the real sphere, is written as 

\begin{equation}
\frac{1}{\epsilon_0} \mathbf{P}_f(\mathbf{r}, \omega)=\int d^3 \mathbf{r'} [\bm{\epsilon}(\mathbf{r}-\mathbf{r'}, \omega)- 
\epsilon_{\infty} \mathbf{I} \delta(\mathbf{r}-\mathbf{r'})] \cdot \mathbf{E}^{M}(\mathbf{r'}, \omega),
\label{P-general}
\end{equation}
where  $\mathbf{E}^{M}$ is the electric field vector in the infinite medium, 
which, for $|\mathbf{r}|< R$ is the actual electric field inside the
sphere, $\epsilon_0$ is the permittivity of free space, $\bm{\epsilon}$ is the dielectric tensor of the medium,
 $\epsilon_{\infty} \mathbf{I}$ accounts for interband transitions of bound electrons, and the integral extends to the whole space. 
In this integral, the region of space  $|\mathbf{r'}|< R$ describes the excitations
produced at a point $\mathbf{r'}$ inside the sphere that propagate directly to the point $\mathbf{r}$ while the fictitious region 
$|\mathbf{r'}|> R$ simulates these excitations that arrive at $\mathbf{r}$ after being reflected at the surface.  
Therefore, the surface properties are mimicked
by the values of the electric field $\mathbf{E}^{M}$ in the fictitious region of the infinite medium.
 In this work, we want the sphere surface to reflect electrons completely at random
and this means that, on the average, no excitation will arrive to $\mathbf{r}$ coming from the surface.
In eq. (\ref{P-general}), this is thus equivalent to making the electric field $\mathbf{E}^{M}$ to be zero outside the sphere.
Hence our problem consists of constructing an electric field of the form
\begin{equation}
\mathbf{E}^{M}(\mathbf{r}, \omega)=\left\{
\begin{array}{ll}
\mathbf{E}_{sphere}(\mathbf{r}, \omega) & \mathrm{for}\;\; |\mathbf{r}|< R\nonumber \\
0 & \mathrm{for}\;\; |\mathbf{r}|> R
\end{array} \right.
\end{equation}
satisfying the Maxwell equations. In general, this is not possible without adding fictitious charges/currents to the 
ficticious region of the medium. 
Here we will use a different, equivalent, procedure. 

From the point of view of an electromagnetic theoretical formulation of both kinds of surface effects discussed above, 
one should note that the radial component $E_r$ 
deviates from its classical counterpart because of the existence of longitudinal modes localized in the surface region, 
which are not present in the classical formulation. 
Consequently, the spill-out effect can only be analyzed if a non-local longitudinal dielectric function, that is, dependent on the spatial coordinates as well as on the frequency, is included in the theory.
In a similar way, a description of the diffuse surface scattering mechanism requires the use of a non-local transverse dielectric function,
which, as we will see below, allows to include in the theory the excitation of transverse electron-holes pairs in addition to
the classical polariton mode propagating with wave vector $k_t=\frac{\omega}{c}\sqrt{\epsilon(\omega)}$.
Since our previous analysis shows that we can expect the spill-out effects to be small, we will neglect them completely 
by making the longitudinal dielectric function of the sphere a local one. However, the transverse dielectric function has to be non-local.
 In our approximation $\bm{\epsilon}$ only depends on spatial coordinates through the difference $\mathbf{r}-\mathbf{r'}$, and
it is convenient to Fourier-transform the permittivity to momentum space. We will use here the simplest possible form for a non-local transverse dielectric function

\begin{equation}
\epsilon_T(\mathbf{k},\omega)=\epsilon_{\infty}-\frac{\omega_p^2}{\omega^2-\Delta^2+i\omega\gamma_b-\beta_T ^2 k^2},
\label{epsilonT}
\end{equation}
where $\mathbf{k}$ is a wave vector, $|\mathbf{k}|=k$ and $\beta_T$ is a constant proportional to the Fermi velocity $v_F$. 
This form of $\epsilon_T$ is obtained when the whole spectrum of electron-hole pairs is substituted by just a single pair.
Thus the pole of eq. (\ref{epsilonT}) describes a transverse electron-hole pair dispersing in energy linearly with the wave vector.
Quantum size effects (QSE) 
are included in the theory by means of the energy gap  $\Delta=\omega_p\frac{R_0}{R}$ \cite{NJP_15_083044_carmina}
with $R_{0}=\sqrt{\frac{3\pi a_0}{4m^{*} k_F}}$ \cite{SovPhysJETP_21_940_gorkov}, $k_F=(3 \pi^2 n_e)^\frac{1}{3}$ being the Fermi wave vector.
In this model, the strength of the QSE with particle size is proportional to $R_0/R$.

The longitudinal and transverse dielectric functions have to be equal in the $\mathbf{k}=0$ limit and therefore a 
good approximation for $\epsilon_L$ is

\begin{equation}
\epsilon_L(\omega)=\epsilon_{\infty}-\frac{\omega_p^2}{\omega^2-\Delta^2+i\omega\gamma_b}.
\label{epsilonL}
\end{equation}

It is convenient to solve for $\mathbf{E}^{M}$ in a basis set of spherical vectors functions:
$\mathbf{l}_{plm}(k,\mathbf{r})$, $\mathbf{m}_{plm}(k,\mathbf{r})$ and  $\mathbf{n}_{plm}(k,\mathbf{r})$ represented by their quantum numbers, $l, m$ and the
parity $p$ (even or odd) with respect to the azimuthal angle. Here $l$ and $m$ only take positive integer values. 
The electric field vector of a plane wave of wave vector $k_m=\frac {\omega}{c} \sqrt{\epsilon_m}$, incident onto the sphere along the 
$\mathbf{x}$ direction, can be expressed with components having $m=1$,\cite{stratton1941} simplyfying the problem.
Then, the field in the extended medium can be written in general as

\begin{equation}
\mathbf{E}^{M}(\mathbf{r})=\sum_{l=1}^{\infty} i^{l} \frac{2l+1}{l(l+1)}\mathbf{E}_{l}^{M}(\mathbf{r}).
\label{E1}
\end{equation}

The field inside the metal sphere is a linear combination of the different normal modes that can be excited
according to the proposed dielectric functions. In our case, there are no longitudinal modes because $\epsilon_L$ is a local dielectric
function and there are two transverse modes (index $j=1,2$): the polariton-like mode 
of wave vector $T_2$ and one transverse electron-hole pair of wave vector $T_1$ (see SI for details). 
Thus, the field inside the extended medium can be written as
\begin{equation}
\mathbf{E}_{l}^{M}(\mathbf{r}) = \left\{
\begin{array}{ll}
\sum_{j=1,2} [ E_{l,j}^{(m)}\mathbf{m}_{ol1}(T_{j}, \mathbf{r})+ & \\ 
\;\;\;\;\;\;\;\;\;\;\;\;+E_{l,j}^{(n)}\mathbf{n}_{el1}(T_{j}, \mathbf{r})] & \mathrm{for} \;\; r< R \\
0 & \mathrm{for} \;\; r>R 
\end{array} \right.
\label{E-modes}
\end{equation}

At this point, the electric field inside the sphere contains four constants, namely, $E_{l,j}^{(n)}$ and $E_{l,j}^{(m)}$ ($j=1,2$) and
the electric field outside the sphere contains two constants for the scattered amplitudes of the even and odd components. 
However, we have four matching conditions for the electromagnetic field at the sphere surface. The missing two equations can be obtained 
by imposing that the free-polarization $\mathbf{P}_f$ calculated from the definition of eq. (\ref{P-general}) with the field
$\mathbf{E}_{l}^{M}(\mathbf{r})$ fulfills the wave equation 
$\nabla \times(\nabla \times \mathbf{E})-(\frac{\omega}{c})^2 \epsilon_{\infty}\mathbf{E}= (\frac{\omega}{c})^2 \frac{1}{\epsilon_0} \mathbf{P_f}$,
inside the sphere (details in the SI). 

The electric field in the medium outside the sphere, has an odd component given by 
\begin{equation}
\mathbf{E}_{o}^{m}(\mathbf{r})=\sum_{l=1}^{\infty} i^{l} \frac{2l+1}{l(l+1)} [\mathbf{m}_{ol1}(k_m, \mathbf{r})+a_{l}\mathbf{m}_{ol1}^{r}(k_m, \mathbf{r})],
\label{E0o}
\end{equation}
and the even component is
\begin{equation}
\mathbf{E}_{e}^{m}(\mathbf{r})=\sum_{l=1}^{\infty} i^{l} \frac{2l+1}{l(l+1)} [-i\mathbf{n}_{el1}(k_m, \mathbf{r})-i b_{l}\mathbf{n}_{el1}^{r}(k_m, \mathbf{r})].
\label{E0e}
\end{equation}

Solving for $a_{l}$ and $b_{l}$ allows us to calculate the absorption cross section
\begin{equation}
\sigma_{abs}\approx-\frac{2\pi}{k_m^2}\sum_{l=1}^{\infty} (2l+1) \mathrm{Re}[a_{l}+b_{l}],
\label{cross-section}
\end{equation}
as scattering can be neglected in all the results we present due to the small size of our particles. In the following calculations we use the values of the parameters appropriate for ZnO nanocrystals in toluene: 
$\epsilon_{\infty}=3.72$, $\epsilon_m=2.5$, $m^{*}=0.28$ and $\gamma_b=$0.1 eV.\cite{ACSNano_8_1065_schimpf} With these values, we use the lower limit of the experimental electron density, $n_e=1 \times 10^{20}$ cm$^{-3}$, to get the experimental surface plasmon energies at the largest radii. Then, $r_s=25.3$, the Fermi velocity $v_F=0.59 \times 10^{6}$ m s$^{-1}$ and $\omega_p=\sqrt { \frac{n_e e^2}{\epsilon_0 m^{*}m_e}}= 0.70$~eV. 

\begin{figure}
\centering
\includegraphics[width=80mm]{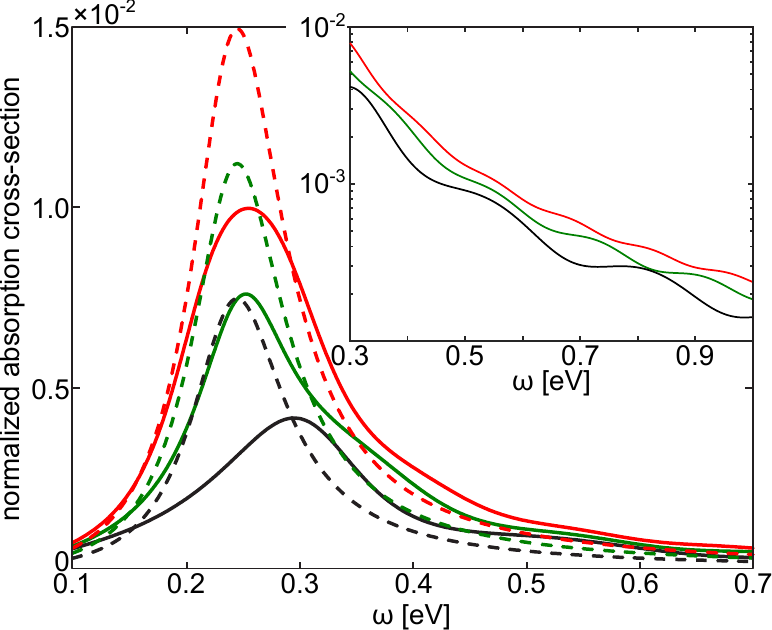}
\caption{Effects of diffuse surface scattering as seen in absorption. Absorption cross-section normalized to surface area as a function of energy for $R=$ 4, 3 and 2 nm from
top to bottom. Quantum size effects are neglected ($\Delta=0$). Dashed lines: classical Mie theory;
continuous lines: present theory for a totally diffuse surface. Notice the asymmetric widening of the resonance and the  strong blue-shift for R = 2 nm.
The inset shows a magnification of the plot in the energy region above the peak (semi-log scale) to highlight the oscillations.}
\label{fig_cross_R432}
\end{figure}

To show clearly the effects a diffuse scattering surface produces in the plasmonic resonances, we first perform a calculation
of the absorption cross section neglecting quantum size effects ($\Delta=0$) and with $\beta_T=v_F/\sqrt 5$. This value of $\beta_T$ is given by the 
long wave length limit of the Lindhard transverse dielectric function.
The results are depicted in Fig. \ref{fig_cross_R432} for spheres of radii
$R=$ 4, 3 and 2 nm, compared to the classical results of the Mie theory \cite{AnnPhys_25_377_mie}. 
We can appreciate the widening of the resonances and the appearance of small oscillations on their right side which are caused by the additional
propagating transverse mode. The resonance strongly blue-shifts in energy  for $R=$ 2 nm. 

From our calculated cross section we extract the position, $\omega_R$, and the width of the resonance, $\Gamma_R$, defined as
the full width at half maximum. $\beta_T$ and $\Delta$ are taken as adjustable parameters given that they have to be increased from the values $\beta_T=v_F/\sqrt 5$ and  
$\Delta=\omega_p\frac{R_0}{R}$, $R_0=\sqrt{\frac{3\pi a_0}{4m^{*} k_F}}=0.56$ nm, by less than 30$\%$
in order to reproduce the experimental results of Ref. \cite{ACSNano_8_1065_schimpf}.
A reason for this is that the value of $\beta_T $ is justified for a large system. Moreover, other values of $\beta_T$ can be
found in the literature since the form of eq. (\ref{epsilonT}) substitutes the whole spectrum of transverse electron-hole pairs by a single pair.
With respect to $\Delta$, a local dielectric function for few electrons in a box has 
been calculated \cite{JPCL_5_3112_prashant}. It has the form of eq. (\ref{epsilonL}) with 
$\Delta_{box}=\pi^{\frac{3}{2}} (\frac{3 N_e}{\pi})^{\frac{1}{3}} \frac{\hbar}{m^{*}a}$, where $N_e$ is the number of electrons 
and $a$ is the side of the box.  Our frequency gap $\Delta$ scales 
with $N_e, m^{*}$ and particle size exactly with the same powers and a prefactor that differs by 15$\%$
if the electron density is the same in both models. Since the prefactor has to depend on the shape of the nanoparticle,
 a change of ca. 20$\%$ is justified.

\begin{figure}
\centering 
\includegraphics[width=80mm]{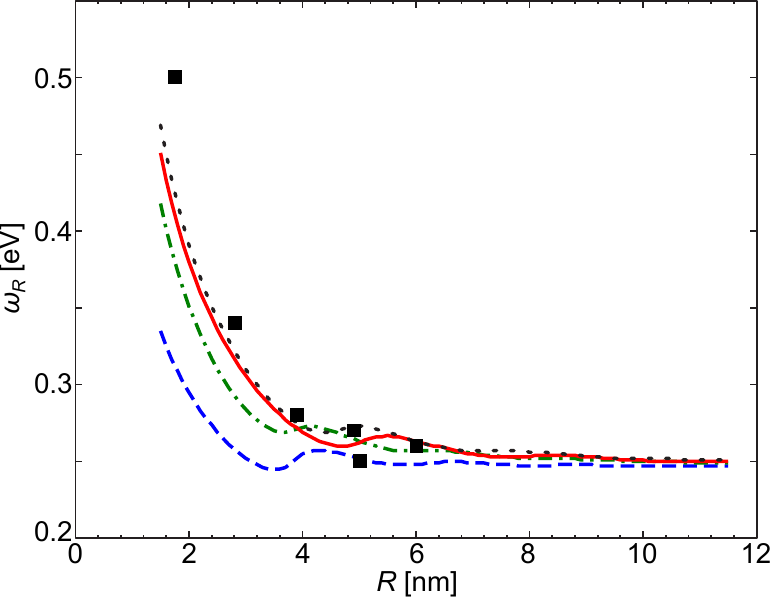}
\caption{The energy of the plasmonic resonance versus the sphere radius $R$ for four groups of the values of the parameters
$\beta_T$ and $R_0$.  Dashed line: $\beta_T=v_F/\sqrt 5$ and $R_0=0$ (no QSE); dot-dashed line $\beta_T=v_F/\sqrt 5$ and $R_0=0.56$ nm;
continuous line: $\beta_T=1.3v_F/\sqrt 5 $ and $R_0=0.56$ nm and dotted line: $\beta_T=1.2v_F/\sqrt 5 $ and $R_0=1.2 \times 0.56$ nm.
 The experimental values of Ref. \cite{ACSNano_8_1065_schimpf} are represented by the black squares.
 By including QSE and slightly increasing the parameters involved from the basic theory there is a good account for the experimental findings.}
\label{fig_wR}
\end{figure}

Figure~\ref{fig_wR} shows $\omega_R$ as a function of the sphere radius. The dashed line 
displays the surface plasmon energy for $\Delta=0$ and $\beta_T=v_F/\sqrt 5$; it presents small oscillations with R 
for radii larger that $\simeq$ 3 nm, and then quickly blue shifts for smaller radii. When we include the quantum size effect using
 $R_0=0.56$ nm (dot-dashed line) we observe the additional blue-shift caused by this effect, 
which is more pronounced for the smaller radii, but the oscillations remain. If we now increase $\beta_T$ by 30$\%$ keeping the same value of $R_0$
(solid line) we further increase the blue-shift at small radii and also change the oscillations at large radii. 
When we increase both, $\beta_T$ and $R_0$ by 20$\%$ we obtain the dotted line which does not differ much from the previous case. 
In both cases the experimental results \cite{ACSNano_8_1065_schimpf} can be reproduced remarkably well.
Thus a diffuse surface produces a non-negligible blue-shift of the plasmonic resonances in small spheres with a low density of electrons,
a result which is not obtained in the simple theory of Ref.\cite{SSComm_52_971_apell} where they would be essentially unshifted.

\begin{figure}
\centering
\includegraphics[width=80mm]{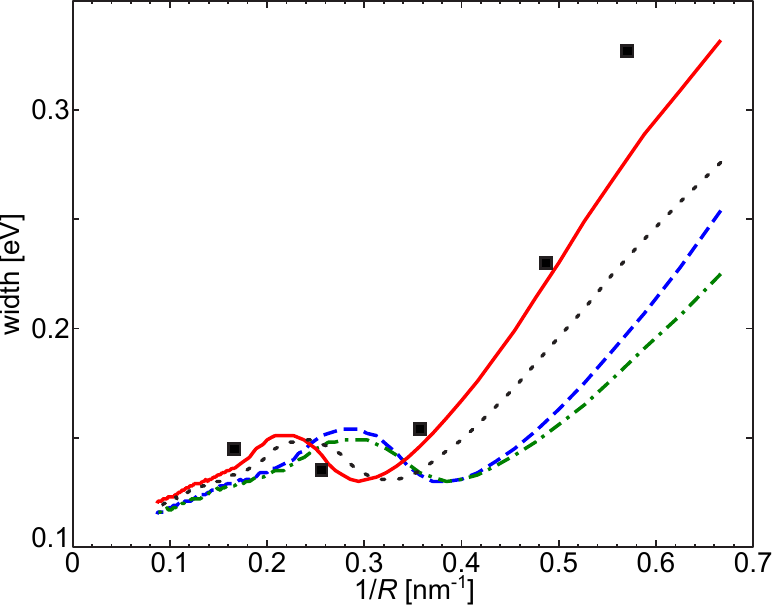}
\caption{The width of the plasmonic resonace versus $1/R$ for the the same groups of the values of the parameters as in Fig. \ref{fig_wR}.
 dashed line: $\beta_T=v_F/\sqrt 5$ and $R_0=0$; dot-dashed line $\beta_T=v_F/\sqrt 5$ and $R_0=0.56$ nm;
continuous line: $\beta_T=1.3v_F/\sqrt 5 $ and $R_0=0.56$ nm and dotted line: $\beta_T=1.2v_F/\sqrt 5 $ and $R_0=1.2 \times 0.56$ nm.
The experimental values of Ref. \cite{ACSNano_8_1065_schimpf} are represented by the black squares. 
While in the simple model of \cite{SSComm_52_971_apell} the plasmon width would increase linearly with $1/R$ for all radii, the detailed model for diffuse surface scattering 
produces this behavior only at  large radii. 
 The width actually oscillates at intermediate radii and then increases fast at  short radii.
Notice that the plasmon  width is much more sensitive to the values  chosen for  $\beta_T$ than the plasmon energy.
The experimental width can be reproduced by the same set of parameters that reproduce the plasmon resonance.}
\label{fig_width}
\end{figure}

Figure \ref{fig_width} depicts the width of the resonance as a function of $1/R$, for the same set of values of $\beta_T$ and $R_0$ as in 
Fig.~\ref{fig_wR}. Note that, while in the
pure classical theory the width of the resonance would be equal to the bulk value $\gamma_b=0.1$ eV (bottom line of Fig.~\ref{fig_width} ), 
the surface scattering mechanism increases the
plasmon damping linearly in $1/R$ for large radii. Then, for 6 nm $ \ge R \ge $ 3 nm the width oscillates between 0.13 eV and 0.16 eV and 
finally, for radii smaller than ca. 3--2.5 nm the 
plasmon width increases again almost linearly with $1/R$ with a higher slope which also increases with $\beta_T$. The
experimental results of Ref. \cite{ACSNano_8_1065_schimpf} are shown by the full squares.  The width is better reproduced by our calculation with the largest
value of $\beta_T$. Note also that the plasmon width is much more sensitive to the values of $\beta_T$ than the plasmon energy.

We now obtain the length $d_{\theta}(\omega)$ from our electromagnetic theory. This can be done analytically given the simplicity
of our transverse dielectric function and that we only need to consider the $l=1$ component of the fields having even symmetry.
Using the small-argument asymptotic values of the spherical Bessel functions appearing in the electric fields of the even mode (see eq. (S36)) we get $d_{\theta}$ as
\begin{equation}
\frac{d_{\theta}(\omega, R)}{R}=-i(\epsilon-\epsilon_{\infty})(tR)h_{1}^{(1)}(tR)j_{1}(tR),
\label{d-theta}
\end{equation}
where $t$ is the pole of the transverse dielectric permittivity, eq. (\ref{epsilonT}), and $\epsilon \equiv \epsilon_T(k=0, \omega)$. 
The origin of the oscillations in $\omega_R$ and $\Gamma_R$ shown in Figs. \ref{fig_wR} and \ref{fig_width}
is clearly related to the Bessel functions appearing in eq. (\ref{d-theta}).  
Since $tR \simeq \frac{\omega R}{\beta_T}$ (eq. (S27)), then $|tR| \gg 1$ for the values of $R$ used here and 
large-argument asymptotic expansions of the Bessel functions can be performed yielding
\begin{equation}
d_{\theta}(\omega, R) \simeq i \frac{\omega_p^2}{(\omega ^2-\Delta^2+i\omega\gamma_b)^{\frac{3}{2}}}
\frac{\beta_T}{2}[1+e^{i 2 t R}].
\label{d-theta-asym}
\end{equation}

Comparing this equation with the corresponding $d_{\parallel}$ for the planar surface, eq. (S13), which is  used in simple models, 
we see that the correction to $d_{\parallel}$ for a real sphere is the factor $1+e^{i 2 t R}$, 
so a large $\mathrm{Im}[t]$ is required to get rid of the oscillatory factor to have the same limit, assuming $\Delta$ is unimportant.
Then the crossover from large values of $R$, where the sphere can still be described by the planar surface, to small values 
of $R$, is given by $ 2 R \simeq (\mathrm{Im}[t])^{-1}$.  
The physical meaning of this relation is clear. The decay length of the transverse electron-hole pair mode is
$(\mathrm{Im}[t])^{-1}$.  When this length becomes shorter than the sphere diameter $2R$, the mode only feels one surface and 
consequently, the limit of the planar surface is reached.

\begin{figure}
\centering
\includegraphics[width=160mm]{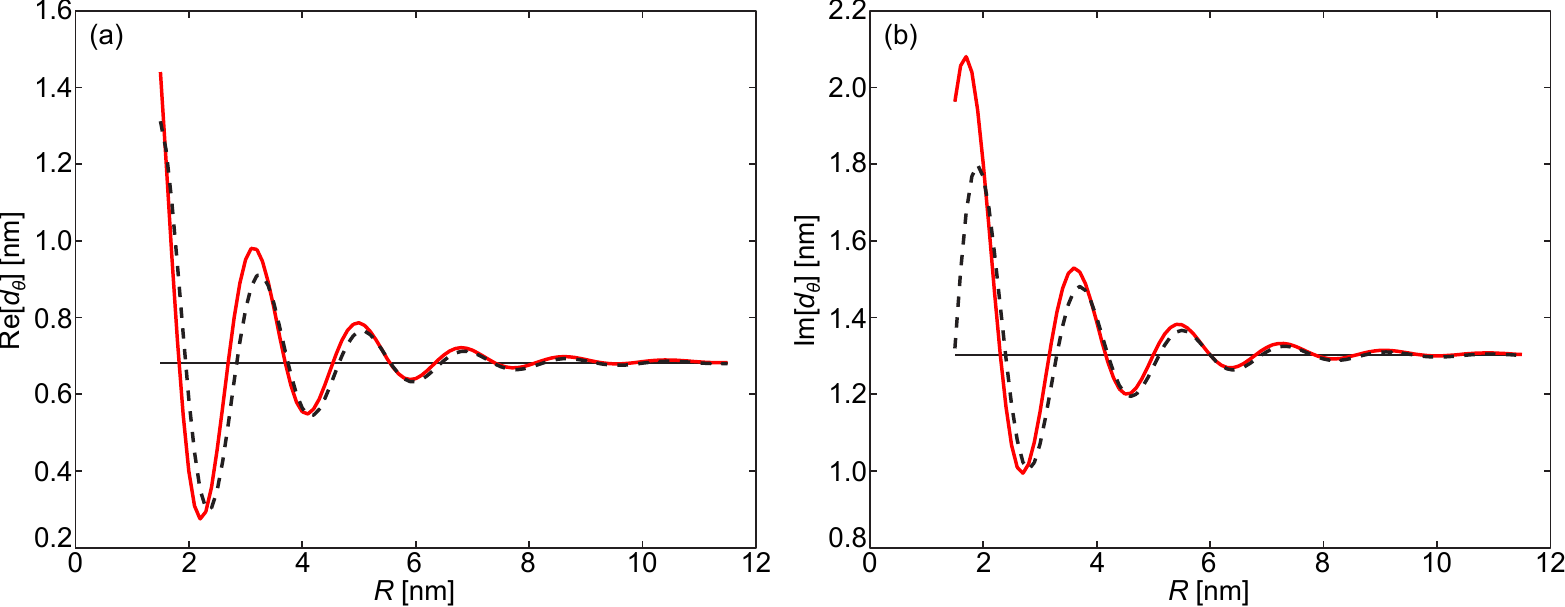}
\caption{(a) The real part and (b) the imaginary part of $d_{\theta}$ versus  $R$ for $\beta_T=v_F/\sqrt 5$, $R_0=0$ and $\omega_t=0.3$ eV.
Red solid line: results of the full theory, eq. (\ref{d-theta}), and dashed line: results of the asymtotic expansion eq. (\ref{d-theta-asym}).
The continuous horizontal lines are the values of $\mathrm{Re}[d_{\parallel}(\omega_t)]$ and $\mathrm{Im}[d_{\parallel}(\omega_t)]$, being good asymptotes for  $d_{\theta}$.}
\label{fig_56_d_theta_vs_R}
\end{figure}

Figure \ref{fig_56_d_theta_vs_R} shows the real and the imaginary parts of $d_{\theta}$ as a function of $R$ 
compared
to their large-argument asymptotic values given by eq. (\ref{d-theta-asym}), for a typical value of $\omega_t=0.3$~eV, $\beta_T=v_F/\sqrt 5$
 and $R_0=0$ (no QSE).
The non-zero values of $\mathrm{Re}[d_{\theta}]$ at large values of $R$ are
the consequence of having  $\frac{\gamma_b}{\omega_t}=\frac{1}{3}$ in this calculation, which, however, produces negligible shifts
of the surface plasmon energies shown in Fig. \ref{fig_wR}. Note that, even if QSE are neglected, 
$d_{\theta}$ depends strongly on $R$ for small radii.
One can note that the expansion of eq. (\ref{d-theta-asym}) is a good approximation to $d_{\theta}$ for radii larger than ca. 2 nm. 

Surface scattering is usually included in simple theories by means of an effective damping rate  $\gamma=\gamma_b+A\frac{v_F}{R}$,
 $A$ being a constant of the order of 1. This procedure produces a linear increase of the LSPR width with $1/R$
and, for low electron density systems, a continuous red shift of its 
 spectral position with decreasing $R$ \cite{ACSNano_8_1065_schimpf}. 
 While this may be a reasonable approximation for large radii, when $Re[d_{\theta}]/R \ll1$ and the main effect
of surface scattering is to increase the plasmon width linearly with $1/R$ while the plasmon energy presents small oscillations around its classical value
(Figs. \ref{fig_width} and \ref{fig_wR} respectively), an effective damping rate theory cannot capture the nature of diffuse surface scattering at short radii.
 As already noted, $d_{\theta}(\omega, R)$ is a complex quantity whose real part cannot in general be neglected. A better approximation can be obtained by defining 
an effective size-dependent dielectric function for a sphere containing diffuse surface scattering, $\tilde\epsilon(\omega, R)$,
via the Rayleigh polarizability (see eq. (S45)) as
\begin{equation}
\tilde\epsilon(\omega, R)=\epsilon+2 \frac{d_{\theta}(\omega, R)}{R},
\end{equation}
and $d_{\theta}(\omega, R)$ expressed by eq. (\ref{d-theta}). Such a definition can be useful in the analysis of problems involving
the dielectric function of the nanosphere.

In this article, we have investigated how different surface effects, namely electronic spill-out and diffuse scattering, impact in the
optical properties of ultra-low electron density nanospheres which are experimentally accessible by electron doping
of semiconductor quantum dots. 
We first estimated the relative contributions of spill-out to diffuse surface scattering, showing that the ratio decreases with effective electron mass 
and with increasing $r_s$. 
Next, we used a more elaborated theoretical model 
for including the diffuse surface scattering mechanism into an electromagnetic theory. Using this model, we calculated the extinction cross-section and 
 found that the plasmonic resonances show an oscillatory behavior both in position and width 
for sizes larger than ca. 3 nm  in radius and a strong blueshift in energy and an increased width for smaller radii. When our model
is used in conjunction with a dielectric function that includes effects of quantization due to the small size of the particles, 
we are able to reproduce the position 
and widths of the measured resonances \cite{ACSNano_8_1065_schimpf} using reasonable values of the 
two basic parameters, $\beta_T$ and $\Delta$, involved in the theory.
From our formalism, we can extract the length $d_{\theta}(\omega,R)$ that allows to include the effects of a diffuse surface 
into an effective size-dependent dielectric function, which can be useful for future work.
We conclude that diffuse scattering at the surface of ultra-low electron density nanocrystals is an important mechanism affecting their plasmonic properties 
and has to be taken into account when designing plasmonic devices based on doped semiconductor nano structures.

\begin{acknowledgement}
RCM acknowledges financial support from the Spanish Mineco via the project MAT2014-53432-C5-5-R. 
TJA thanks the Foundation of Polish Science for support via the project HOMING PLUS/2013-7/1. 
TJA and SPA acknowledge financial support from the Swedish Foundation for Strategic Research via the 
Functional Electromagnetic Metamaterials for Optical Sensing project SSF~RMA~11.
\end{acknowledgement}

\begin{suppinfo}
Details of the simple theoretical model and estimation of spill-out and diffuse surface scattering effects.
Details of the solution for the electromagnetic fields in the more detailed theory for diffuse surface scattering.
Derivation of the formulae for the diffuse surface scattering length $d_{\theta}(\omega, R)$ and the 
corresponding size-dependent dielectric function.
\end{suppinfo}

 
\providecommand{\latin}[1]{#1}
\providecommand*\mcitethebibliography{\thebibliography}
\csname @ifundefined\endcsname{endmcitethebibliography}
  {\let\endmcitethebibliography\endthebibliography}{}

\newpage

\end{document}